\documentclass[11pt]{article}
\usepackage{epsf}

\begin{document}


\newtheorem{theorem}{Theorem}[section]
\newtheorem{itlemma}{Lemma}[section]
\newtheorem{itdefinition}{Definition}[section]
\newtheorem{itexample}{Example}
\newtheorem{itclaim}{Claim}[section]
\newtheorem{itproposition}{Proposition}[section]
\newtheorem{itremark}{Remark}[section]
\newtheorem{itcorollary}{Corollary}[section]

\newenvironment{example}{\begin{itexample}\rm}{\end{itexample}}
\newenvironment{definition}{\begin{itdefinition}\rm}{\end{itdefinition}}
\newenvironment{lemma}{\begin{itlemma}\rm}{\end{itlemma}}
\newenvironment{corollary}{\begin{itcorollary}\rm}{\end{itcorollary}}
\newenvironment{claim}{\begin{itclaim}\rm}{\end{itclaim}}
\newenvironment{proposition}{\begin{itproposition}\rm}{\end{itproposition}}
\newenvironment{remark}{\begin{itremark}\rm}{\end{itremark}}

\newcommand{\qed}{\hfill \halmos} 
\newcommand{\mybox}{\hfill $\Box$} 

\newcommand{\comment}[1]{}
\newcommand{\halmos}{\rule{1ex}{1.4ex}}
\newenvironment{proof}{\noindent {\em Proof}.\ }{\hspace*{\fill}$\halmos$
\medskip}

\def\vbar{\mathchoice{\vrule height6.3ptdepth-.5ptwidth.8pt\kern-.8pt}
   {\vrule height6.3ptdepth-.5ptwidth.8pt\kern-.8pt}
   {\vrule height4.1ptdepth-.35ptwidth.6pt\kern-.6pt}
   {\vrule height3.1ptdepth-.25ptwidth.5pt\kern-.5pt}}
\def\fudge{\mathchoice{}{}{\mkern.5mu}{\mkern.8mu}}
\def\bbc#1#2{{\rm \mkern#2mu\vbar\mkern-#2mu#1}}
\def\bbb#1{{\rm I\mkern-3.5mu #1}}
\def\bba#1#2{{\rm #1\mkern-#2mu\fudge #1}}
\def\bb#1{{\count4=`#1 \advance\count4by-64 \ifcase\count4\or\bba A{11.5}\or
   \bbb B\or\bbc C{5}\or\bbb D\or\bbb E\or\bbb F \or\bbc G{5}\or\bbb H\or
   \bbb I\or\bbc J{3}\or\bbb K\or\bbb L \or\bbb M\or\bbb N\or\bbc O{5} \or
   \bbb P\or\bbc Q{5}\or\bbb R\or\bbc S{4.2}\or\bba T{10.5}\or\bbc U{5}\or
   \bba V{12}\or\bba W{16.5}\or\bba X{11}\or\bba Y{11.7}\or\bba Z{7.5}\fi}}

\def\Q{{\bb Q}}                         
\def\N{{\bb N}}                         
\def\R{{\bb R}}                         
\def\I{{\bb Z}}                         
\def\B{{\bb B}}                         

\def\rmtr{{\rm tr}}

\def\nasymptotic{{_{\stackrel{\displaystyle\longrightarrow}
{N\rightarrow\infty}}\,\, }} 
\def\masymptotic{{_{\stackrel{\displaystyle\longrightarrow}
{M\rightarrow\infty}}\,\, }} 
\def\wasymptotic{{_{\stackrel{\displaystyle\longrightarrow}
{w\rightarrow\infty}}\,\, }} 

\def\asymptext{\raisebox{.6ex}{${_{\stackrel{\displaystyle\longrightarro
w}{x\rightarrow\pm\infty}}\,\, }$}} 
\def\epsilim{{_{\textstyle{\rm lim}}\atop_{\epsilon\rightarrow 0+}\,\, }}

\def\beqra{\begin{eqnarray}} \def\eeqra{\end{eqnarray}}
\def\beqast{\begin{eqnarray*}} \def\eeqast{\end{eqnarray*}}
\def\beq{\begin{equation}}      \def\eeq{\end{equation}}
\def\be{\begin{enumerate}}   \def\ee{\end{enumerate}}

\def\bet{\beta}
\def\gam{\gamma}
\def\Gam{\Gamma}
\def\la{\lambda}
\def\eps{\epsilon}
\def\La{\Lambda}
\def\si{\sigma}
\def\Si{\Sigma}
\def\al{\alpha}
\def\Tha{\Theta}
\def\tha{\theta}
\def\vphi{\varphi}
\def\del{\delta}
\def\Del{\Delta}
\def\ab{\alpha\beta}
\def\om{\omega}
\def\Om{\Omega}
\def\mn{\mu\nu}
\def\mun{^{\mu}{}_{\nu}}
\def\kap{\kappa}
\def\rsi{\rho\sigma}
\def\beal{\beta\alpha}

\def\til{\tilde}
\def\rta{\rightarrow}
\def\eqv{\equiv}
\def\nab{\nabla}
\def\pa{\partial}
\def\sit{\tilde\sigma}
\def\ul{\underline}
\def\indt{\parindent2.5em}
\def\nd{\noindent}
\def\var{{1\over 2\si^2}}
\def\ivar{\left(2\pi\si^2\right)}
\def\iivar{\left({1\over 2\pi\si^2}\right)}
\def\caa{{\cal A}}
\def\cb{{\cal B}}
\def\cac{{\cal C}}
\def\cd{{\cal D}}
\def\ce{{\cal E}}
\def\cf{{\cal F}}
\def\cg{{\cal G}}
\def\ch{{\cal H}}
\def\ci{{\cal I}}
\def\cj{{\cal{J}}}
\def\ck{{\cal K}}
\def\cl{{\cal L}}
\def\cm{{\cal M}}
\def\cn{{\cal N}}
\def\cO{{\cal O}}
\def\cp{{\cal P}}
\def\cq{{\cal Q}}
\def\car{{\cal R}}
\def\cs{{\cal S}}
\def\ct{{\cal{T}}}
\def\cu{{\cal{U}}}
\def\cv{{\cal{V}}}
\def\cw{{\cal{W}}}
\def\cx{{\cal{X}}}
\def\cy{{\cal{Y}}}
\def\cz{{\cal{Z}}}

%
\vspace*{-0.5in}
\begin{center}
{\large\bf
Scaling and Universality of the Complexity of Analog Computation}
\end{center}
\vspace{-.2in}
\begin{center}
{\bf Yaniv Avizrats $^{a}$, Joshua Feinberg$^{a,b}$, Shmuel Fishman$^{a}$}
\end{center}
\begin{center}
$^{a)}${Physics Department,}\\
{Technion, Israel Institute of Technology, Haifa 32000, Israel}\\
$^{b)}${Physics Department,}\\
{University of Haifa at Oranim, Tivon 36006, Israel}\\
\end{center}

\begin{minipage}{5.1in}
{\abstract We apply a probabilistic approach to study the computational 
complexity of analog computers which solve linear programming problems. 
We analyze numerically various ensembles of linear programming 
problems and obtain, for each of these ensembles, the probability distribution 
functions of certain quantities which measure the computational complexity, 
known as the convergence rate, the barrier and the computation time. 
We find that in the limit of very large problems these probability 
distributions are universal scaling functions. In other words, the probability 
distribution function for each of these three quantities becomes, in the
limit of large problem size, a function of a single scaling variable, which is 
a certain composition of the quantity in question and the size of the system. 
Moreover, various ensembles studied seem to lead essentially to the same 
scaling functions, which depend only on the variance of the ensemble. 
These results extend analytical and numerical results obtained recently 
for the Gaussian ensemble, and support the conjecture that these scaling 
functions are universal.}
\end{minipage}

\vspace{10pt}


\section{Introduction}
Digital computers are part of our present civilization. There are, however, 
other devices that are capable of computation. These are analog computers, 
that  are ubiquitious computational tools. The most relevant 
examples of analog computers are VLSI devices implementing neural 
networks \cite{Hertznn-optimwang}, or neuromorphic systems \cite{mead}, 
whose structure is directly motivated by the workings of the brain. Various 
processes taking place in living cells can be considered as analog 
computation \cite{bio} as well.

An analog computer is essentially a physical device that performs 
computation, evolving in continuous time and phase space. It is useful to 
model its evolution in phase space by dynamical systems (DS) \cite{ott}, the 
way classical systems such as particles moving in a potential (or electric 
circuits), are modeled. For example, there are dynamical systems (described by
ordinary differential equations) that are used to solve computational 
problems \cite{brockett,faybusovich,helmke-moore}.
This description makes a large set of analytical tools 
and physical intuition, developed for dynamical systems, applicable to the 
analysis of analog computers.

In contrast, the evolution of a digital computer is discrete both in its 
phase space and in time. 
Consequently, the standard theory of computation and computational 
complexity \cite{Papadimitriou} deals with computation in discrete time
and phase space, and is inadequate for the description of
analog computers. The analysis of computation by analog
devices requiers a theory that is valid in continuous time and phase
space.

Since the systems in question are physical systems, the computation time is 
the time required for a system to reach the vicinity of an attractor 
(a stable fixed point in the present work) combined with the time
required to verify that it indeed reached this vicinity.
This time is the elapsed time measured by a clock, contrary
to standard computation theory, where it is the number of discrete 
steps.

In the exploration of physical systems, it is sometimes much easier to study
statistical ensembles of systems, estimating their typical behavior using
statistical methods \cite{rmt,nucl,chaos}. 
In \cite{bffs, pla} a statistical theory was used to calculate the 
computational complexity of a standard representative problem, namely Linear 
Programming (LP), as solved by a DS.

A framework for computing with DS that converge exponentially to fixed
points was proposed in \cite{dds}. For such systems it is natural to
consider the {\em attracting fixed point as the output}. The
input can be modeled in various ways. One possible choice is
the initial condition. This is appropriate when the aim of
the computation is to decide to which attractor out of many
possible ones the system flows \cite{SF}. Here, in \cite{bffs, pla}, as well 
as in \cite{dds}, the parameters on which the DS depends
(e.g., the parameters appearing in the vector field $F$ in
(\ref{contsys})) are the input.

The basic entity of the computational model is a dynamical
system \cite{ott}, that may be defined by a set of
Ordinary Differential Equations (ODEs)
\beq\label{ode}
\frac{dx}{dt}=F(x),
\label{contsys}
\eeq
where $x$ is an $n$-dimensional vector, and $F$ is an
$n$-dimensional smooth vector field, which converges
exponentially to a fixed point. Eq. (\ref{contsys})
solves a computational problem as follows: Given an instance
of the problem, the parameters of the vector field $F$ are
set (i.e., the input is read), and it is started from some pre-determined 
initial condition. The result of the computation is then deduced
from the fixed point that the system approaches.

In our model we assume we have a physical implementation of the flow 
equation (\ref{contsys}). Thus, the vector field $F$ need not be computed, 
and the computation time is determined by the convergence time to the 
attractive fixed point. In other words, the time of flow to the vicinity
of the attractor is a good measure of complexity, namely the computational 
effort, for the class of continuous dynamical systems introduced above 
\cite{dds}.

In this paper, as in \cite{bffs, pla}, we will study a specific algorithm 
for the solution of the LP problem \cite{toda}. 
We will consider real-continuous
inputs, as the ones found in physical experiments, and that are studied in 
the BSS model \cite{BCSS}, as well as integer-valued inputs. 
For computational models defined on the real numbers, worst case behavior, 
that is traditionally studied in computer science, can be ill-defined and 
lead to infinite computation times, in particular, for some methods
for solving LP \cite{BCSS, traub}. Therefore, we compute the
distribution of computation times for a probabilistic model
of LP instances with various distributions of the data like
in \cite{lpsmaleTodd-models,shamir}. Ill-defined instances constitute a set 
of measure zero in our countinuous probability ensembles, and need not be 
concerned about. In the discrete probability ensembles, we treat them by
appropriate regularization.

The computational complexity of the method presented in \cite{bffs, pla} 
and discussed here is $\cO(n\log n)$, compared to $\cO(n^{3.5} \log n)$ found 
for standard interior point methods \cite{Ye-book}. The basic
reason is that for standard methods (such as interior point methods), the 
major component of the complexity of each iteration is $\cO(n^3)$ due to 
matrix decomposition and inversion of the constraint matrix, while here, 
because of its analog nature, the system just flows according to its 
equations of motion (which need not be computed).

Since we consider the evolution of a {\em vector field}, our model is
inherently parallel. Therefore, to make the analog vs. digital comparison
entirely fair, we should compare the complexity of our method to that of the 
best parallel algorithm. The latter can reduce the $\cO(n^3)$ time needed for 
matrix decomposition/inversion to polylogarithmic time (for well-posed 
problems), at the cost of $\cO(n^{2.5})$ processors \cite{reif}, while our 
system of equations (\ref{contsys}) uses only $\cO(n)$ variables.

The main result of \cite{bffs, pla}, in which LP problems were drawn from 
the Gaussian distribution of the parameters of $F$ (namely, the 
constraints and cost function in (\ref{standard})), was that the 
distribution functions of various
quantities that characterize the computational complexity, were found to be 
scaling functions in the limit of LP problems of large size. In particular,
it was found that these distribution functions depend on the various 
parameters only via specific combinations, namely, the scaling variables.
Such behavior is analogous to the situation found for the
central limit theorem, for critical phenomena \cite{wilson}
and for Anderson localization \cite{anderson}, in spite of the very different 
nature of these problems. It was demonstrated in \cite{bffs, pla} how for 
the implementation of the LP problem on a physical device, methods used in
theoretical physics enable to describe the distribution
of computation times in a simple and physically transparent form.
Based on experience with certain universality properties of
rectangular and chiral random matrix models \cite{universality},
it was conjectured in \cite{bffs, pla} that some universality for 
computational problems should be expected and should be explored. That is, 
the scaling properties that were found for the Gaussian distributions should 
hold also for other distributions. In particular, some specific 
questions were raised in \cite{bffs, pla}:
Is the Gaussian nature of the ensemble unimportant in analogy with
\cite{universality}? Are there universality classes \cite{wilson}
of analog computational problems, and if they exist, what are they?

Thus, we extend the earlier analysis \cite{bffs, pla} of the Gaussian 
distribution to other probability distributions of LP problems, and 
demonstrate {\em numerically} that the distribution functions of various 
quantities that characterize the computational complexity of the analog 
computer which solves LP problems are indeed {\em universal} 
scaling functions, in the limit of large systems. These universal
functions depend upon the original probability ensemble of inputs only via 
the scaling variables, that are proportional to the ones found for the 
Gaussian distribution. For some distributions
of LP problems the scaling variables are even identical (not just 
proportional) to the ones found for the Gaussian distribution. For 
other distributions, on the other hand, where some of the parameters defining
an LP problem may vanish at random (the so-called diluted ensembles in Section
4), either the convergence to universality is much slower, or universality 
is only approximate.

The distribution of constraints and cost function of the LP problems 
that are used in practice is not known. Therefore, the universality of the 
distribution functions of the computation time and other quantities related
to computational complexity is of great importance. It would imply that it may
hold also for the distributions of the LP problems solved in applications. In 
this paper we demonstrate numerically that for several probability 
distributions universality is satisfied, providing support for the conjecture 
that it holds in general.

This paper is organized as follows: In the next section we briefly review 
the dynamical system which solves LP problems. In section 3 we summarize 
the scaling results of \cite{bffs, pla} for the Gaussian ensemble. In section
4 we present our numerical results for the distribution functions of various 
quantities that characterize the computational complexity of the analog 
computer for non-Gaussian probability ensembles. In section 5 we 
demonstrate that these distributions are indeed universal scaling functions. 
Finally, in section 6 we discuss the significance of our results and also 
pose some open problems.

\newpage
\section{A dynamical flow for linear programming}
Linear programming is a P-complete problem \cite{Papadimitriou}, i.e.
it is representative of all problems that can be solved in polynomial time.
The {\it standard form} of LP is
to find \begin{equation}
\label{standard}
\max \{ c^{T}x~:~ x\in \R ^{n},
  A x =  b,x \geq 0  \}
\end{equation}
where $c \in \R^n, b \in \R^m,
A \in \R^{m \times n}$
and $m\leq n$.
The set generated by the constraints in (\ref{standard}) is a polyheder.
If a bounded optimal solution exists, it is obtained at one of its vertices.
The vector defining this optimal vertex can be decomposed
(in an appropriate basis) in the form
$x=(x_{{\cal N}},x_{{\cal B}})$
where $x_{\cal N} = 0$ is an $n-m$ component
vector, while $x_{\cal B} =  B^{-1}  b \geq  0$
is an $m$ component vector, and $B$ is the $m \times m$ matrix whose
columns are
the columns of $A$ with indices identical to the ones of $x_{\cal B}$.
Similarly, we decompose $A=(N,B)$.

A flow of the form (\ref{contsys}) converging to the optimal vertex,
introduced by Faybusovich \cite{faybusovich} will be studied here.
Its vector field $F$ is a projection of the gradient of the
cost function $c^T x$ onto the constraint set, relative to a
Riemannian metric which enforces the positivity constraints $x\geq 0$
\cite{faybusovich}. It is given by
\begin{equation} \label{field}
	F(x)=[X - X A^{T}
		(A X A^{T})^{-1} A X]\: c\; ,
\end{equation}
where $X$ is the diagonal matrix $\mbox{Diag}(x_1 \dots x_n)$.
The $nm + n$ entries of $A$ and $c$, namely, the parameters of the 
vector field $F$, constitute the input; as in other models of 
computation, we ignore the time it takes to ``load'' the input, since this 
step does not reflect the complexity of the computation being performed, 
either in analog or digital computation.
It was shown in \cite{toda} that the flow equations 
given by
(\ref{contsys}) and (\ref{field}) are, in fact, part of a system of
Hamiltonian equations of motion of a completely integrable system of a
Toda type. Therefore, like the Toda system, it is integrable with the
formal solution \cite{faybusovich}
\begin{equation}
\label{solution}
x_i(t) = x_i(0) \exp \left( -\Delta_i t +
\sum_{j=1}^{m} \alpha_{ji} \log \frac{x_{j+n-m}(t)}{x_{j+n-m}(0)} \right)
\end{equation}
($i = 1,\ldots ,n-m$), that describes the time evolution of the $n-m$
independent variables $x_{\cal N}(t)$, in terms of the variables
$x_{\cal B}(t)$. In (\ref{solution})
$x_i(0)$ and $x_{j+n-m}(0)$ are components of the initial
condition, $x_{j+n-m}(t)$ are the $x_{\cb}$ components of the solution,
$\alpha_{ji}= - (B^{-1} N)_{ji}$
is an $m \times (n-m)$ matrix, while
\begin{equation}\label{deltas}
\Delta_i = -c_i  - \sum_{j=1}^{m} c_j \alpha_{ji}\,.
\end{equation}
For the decomposition
$x=(x_{{\cal N}},x_{{\cal B}})$
used for the optimal vertex $\Delta_i \geq 0~~i=1,\ldots,n-m\,,$
and $x_{\cal N}(t)$ converges to 0, while
$x_{\cal B}(t)$ converges to $x^*=B^{-1}b$.
Note that the analytical solution is only a {\em formal} one, and does not
provide an answer
to the LP instance, since the $\Delta_i$ depend on the partition of $A$, and
only relative to a partition corresponding to a
maximum vertex are all the $\Delta_i$ positive.

The second term in the exponent in (\ref{solution}),
when it is positive, is a kind of ``barrier'':
$\Delta_{i}t$
must be larger than the barrier before $x_i$ can decrease to zero.
In the following we ignore the contribution of the initial condition
and denote the value of this term in the infinite time limit by
\begin{equation}
\label{barrier}
\beta_i = \sum_{j=1}^m \alpha_{ji} \log x_{j+n-m}^*.
\end{equation}
Note that although one (or more) of the $x_{j+n-m}^*$ may vanish, in the 
probabilistic ensemble studied here, such an event is of measure zero for 
the continuous probability distributions, as well as for the regularized 
discretized ones (see (\ref{regularization})), and therefore should not be 
considered.
In order for $x(t)$ to be close to the maximum vertex we must have
$x_i(t) < \epsilon$ for $i=1,\ldots,n-m$ for some small positive
$\epsilon$, namely
$\exp (- \Delta_i t + \beta_i) < \epsilon ~,~~ \mbox{for}~ i =
1,\ldots,n-m.$
Therefore we consider
\begin{equation}\label{T}
T = \max_{i} \left( \frac{\beta_i}{\Delta_i} +
    \frac{|\log \epsilon |}{\Delta_i} \right)~,
\end{equation}
as the computation time.
We denote
\begin{equation}\label{Deltamin}
\Delta_{\min} = \min_i \Delta_i,~~~\beta_{\max} = \max_i \beta_i \;.
\end{equation}
The $\Delta_i$ can be arbitrarily small when the inputs are real numbers,
but in the probabilistic model,
``bad'' instances, resulting in computation taking arbitrarily long time, 
are rare as is clear from\footnote{Strictly speaking, 
(\ref{scaling.delta}) was derived in \cite{bffs, pla} for a probabilistic 
model in which the components of $(A,b,c)$ were independent identically 
distributed Gaussian random variables. However, one of the main points of 
this paper is that (\ref{scaling.delta}) is valid for a broad class of 
probability distributions of LP problems.} (\ref{scaling.delta}).

\newpage
\section{Probability distributions and scaling}

Consider an ensemble of LP problems in which the components of $(A,b,c)$ are 
independent identically distributed (i.i.d.) random variables taken from
various {\em even} distributions, with 0 mean and bounded variance.
In a probabilistic model of LP instances, $\Delta_{\min},\, \beta_{\max}$ 
and $T$ are random variables.
Since the expression for $\Delta_i$, equation (\ref{deltas}),
is independent of $b$, its distribution is independent of $b$.
For a given realization of $A$ and $c$, with a partition of $A$
into $(N,B)$ such that $\Delta_i \geq 0$, there exists\footnote{The existence
of $b$ is guaranteed by the fact that the various probability distributions 
are even. See \cite{bffs}, Lemma 3.1. The latter was proved in \cite{bffs} 
under the assumption that the components of $(A,b,c)$ were i.i.d. Gaussian 
variables, but the proof extends trivially to the class of probability
ensembles of the type specified above.}
a vector $b$ such that the resulting polyheder has a bounded optimal solution.
Since $b$ in our probabilistic model is independent of $A$ we obtain:
\newline
${\cal P}(\Delta_{\min} < \Delta | \Delta_{\min} > 0,
\mbox{LP instance has a bounded maximum vertex}) =
{\cal P}(\Delta_{\min} < \Delta | \Delta_{\min} > 0)$.

In \cite{bffs, pla}, the components of $A$, $b$, and $c$ 
were taken from the Gaussian distribution (see, e.g., Eqs.(12-18) 
in \cite{bffs}) with zero mean and variance $\sigma^2$, that was taken as 
unity in the numerical calculations. It was found analytically, in the large 
$(n,m)$ limit, that the probability ${\cal P}(\Delta_{min} < \Delta|
\Delta_{min} > 0) \equiv {\cal F}^{(n,m)}(\Delta)$
is of the scaling form
\begin{equation} \label{scaling.delta}
{\cal F}^{(n,m)}(\Delta)=1-e^{x_\Delta^2}\,{\rm erfc}(x_\Delta)\ \equiv {\cal
F}(x_\Delta) .
\end{equation}
with the scaling variable 
\begin{equation} \label{scalingvariable}
x_\Delta (n,m) =  {1\over\sqrt{\pi}}\left({n\over m}-1\right)\,
{\sqrt{m} \Delta\over\sigma}\, .
\end{equation}
The scaling function ${\cal F}$ contains {\em all} asymptotic information on 
$\Delta$. The probability density function derived from 
${\cal F}(x_\Delta)$ is very wide and does not have a finite variance. 
Also the average of $1/x_\Delta$ diverges.

The amazing point is that in the limit of large $m$ and $n$, the probability 
distribution of $\Delta_{\rm min}$ depends on the variables $m$, $n$ and 
$\Delta$ only via the scaling variable $x_\Delta$. For future reference 
it is convenient to write $x_\Delta$ in the form
\begin{equation}
x_\Delta = a_\Delta^{(g)}  (n/m) \sqrt{m} \Delta\,,
\label{xd}
\end{equation}
with
\begin{equation}
a_\Delta^{(g)}(n/m)  = \frac{1}{\sqrt{\pi}} \left(\frac{n}{m}-1\right) 
\frac{1}{\sigma}\,,              
\label{ad}
\end{equation}
where the superscript refers to the Gaussian distribution. 
If the limit of infinite $m$ and $n$ is taken, so that $n/m$ is fixed, 
$a_\Delta^{(g)} $ is constant. It was verified numerically that for the 
Gaussian ensemble (\ref{scaling.delta}) was a good approximation already for 
$m=20$, and $n=40$.

The existence of scaling functions like (\ref{scaling.delta}) for the
barrier $\beta_{max}$, that is the maximum of the $\beta_i$ defined by 
(\ref{barrier}) and for  $T$ defined by (\ref{T}) (assuming that $\epsilon$ 
is not too small so that the first term in (\ref{T}) dominates) was verified 
numerically for the Gaussian distribution \cite{bffs, pla}. 
In particular for fixed $m/n$, we found that
\begin{equation} \label{scaling.beta}
{\cal P}(1/\beta_{max}~<~1/\beta) =  {\cal
F}^{(n,m)}_{1/\beta_{max}}(1/\beta)\equiv {\cal F}_{1/\beta}(x_\beta)
\end{equation}
and
\begin{equation} \label{scaling.T}
{\cal P}(1/T~<~1/t) = {\cal
F}^{(n,m)}_{1/T}(1/t)\equiv {\cal F}_{1/T}(x_T).
\end{equation}
The corresponding scaling variables are 
\begin{equation}\label{xbgauss}
x_\beta = a_\beta^{(g)} (n/m)\,{m\over\beta}
\end{equation}
and
\begin{equation}\label{xTgauss}
x_T = a_T^{(g)} (n/ m)\,{m\log m\over t}\,.
\end{equation}
Since the distribution functions (\ref{scaling.beta}) and (\ref{scaling.T}) 
are not known analytically, and since $n=2m$ was taken in the numerical 
investigations in \cite{bffs}, we can set arbitrarily 
$a_\beta^{(g)}(2)=a_T^{(g)}(2)=1$.

The scaling functions (\ref{scaling.delta}),  (\ref{scaling.beta}) and
(\ref{scaling.T})
imply the asymptotic behavior
\begin{eqnarray}\label{om}
 1/\Delta_{\min} \sim \sqrt{m}, ~~~~
\beta_{\max} \sim m ,~~~
t \sim m \log m
\end{eqnarray}
with ``high probability'' \cite{bffs, pla}.

\newpage
\section{Non-Gaussian distributions}
In this section we present the results of our numerical calculations of 
the distribution functions of $\Delta_{min}$, $1/\beta_{max}$, and $1/T$ for 
various probability distributions of $A$, $b$, and $c$. 
For this purpose we generated full LP instances $(A,b,c)$ with the probability 
distribution in question. For each instance the LP problem was solved using 
the linear programming solver of MatLab. Only instances with a bounded optimal 
solution were kept, and $\Delta_{\min}$ was computed relative to the optimal 
partition and optimality was verified by checking that $\Delta_{\min} >0$.
Using the sampled instances we obtain an estimate of ${\cal F}^{(n,m)}(\Delta)
={\cal P}(\Delta_{min} < \Delta |\Delta_{min} > 0)$, and of the corresponding 
cumulative distribution functions of the barrier $\beta_{\max}$ and the 
computation time T.

The solution of the  LP problem is used here in order to identify the optimal 
partition of $A$ into $B$ and $N$. This enables one to compute 
$\Delta_{min}$, $\beta_{max}$, and $t$ from (\ref{deltas}), (\ref{barrier}), 
(\ref{T}) and (\ref{Deltamin}), and the distributions 
\begin{equation}
P_\Delta(\Delta) = {\cal P}(\Delta_{min}< \Delta | \Delta_{min}>0)\,,         
\label{pd}
\end{equation}
\begin{equation}
P_\beta(1/\beta) = {\cal P}(1/\beta_{max}<1/\beta)                      
\label{pb}
\end{equation}
and
\begin{equation}
P_T(1/t) = {\cal P}(1/T<1/t)
\label{pt}
\end{equation}
are obtained.

It is convenient at first to keep $n/m$ fixed\footnote{In fact, in all our 
numerical simulations we kept $n/m=2$ fixed.} and to compute these 
distributions as functions of the corresponding scaling variables
\begin{equation}
x'_\Delta = \sqrt{m}\Delta                                    
\label{xd'}
\end{equation}
\begin{equation}
x'_\beta = m/\beta                                          
\label{xb'}
\end{equation}
and
\begin{equation}
x'_T = m \log {m}/t
\label{xt'}
\end{equation}
These are proportional to $x_\Delta, x_\beta$ and $x_T$ defined by 
(\ref{scalingvariable}), (\ref{xbgauss}) and 
(\ref{xTgauss}). We turn now to explore in some detail, various distributions.

\subsection{The bimodal distribution}
\begin{figure}
\epsfxsize=8cm
\centerline{\epsffile[60 270 550 610]{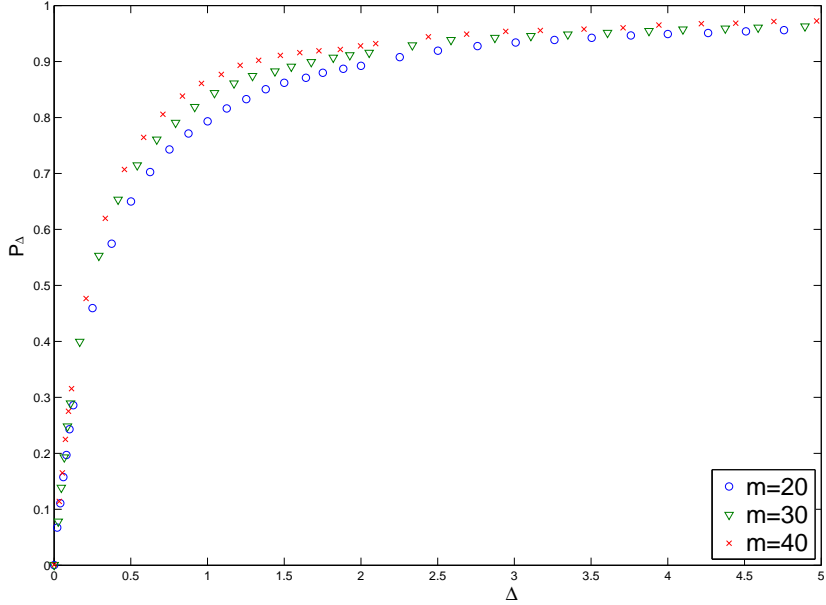}}
\caption{\label{1a}
\ensuremath{ {\cal P}_\Delta }
is plotted as a function of \ensuremath{\Delta}, for the bimodal 
distribution (for \ensuremath{n=2m}). The number of instances 
used in the simulation is 39732 for the $m=20$ case, 46583 for the $m=30$ 
case and 47169 for the $m=40$ case. The number of converging instances for 
each case is 5000.}
\end{figure}

\begin{figure}
\epsfxsize=8cm
\centerline{\epsffile[60 270 550 610]{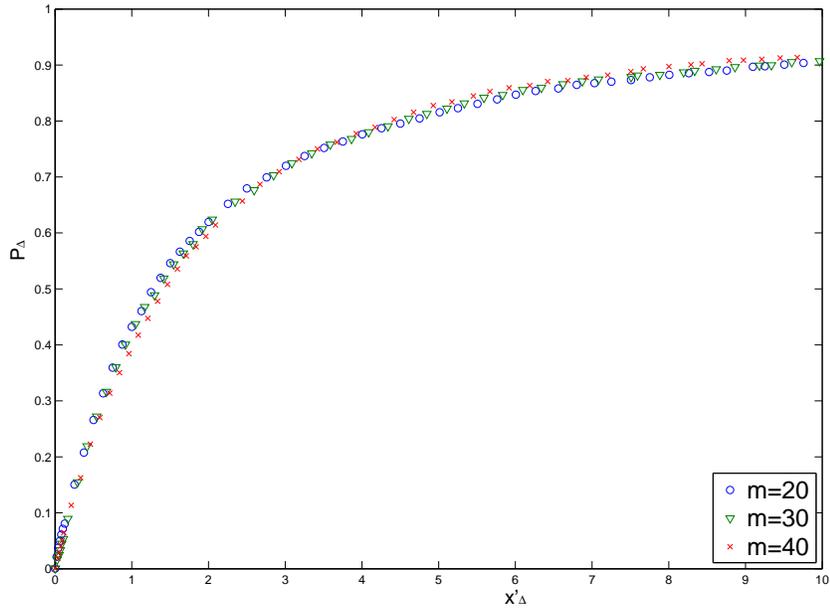}}
\caption{\label{1b}
\ensuremath{ {\cal P}_\Delta }
is plotted as a function of \ensuremath{x'_\Delta} for the bimodal 
distribution for the instances of Fig. \ref{1a}}
\end{figure}

\begin{figure}
\epsfxsize=10cm
\centerline{\epsffile[60 270 550 610]{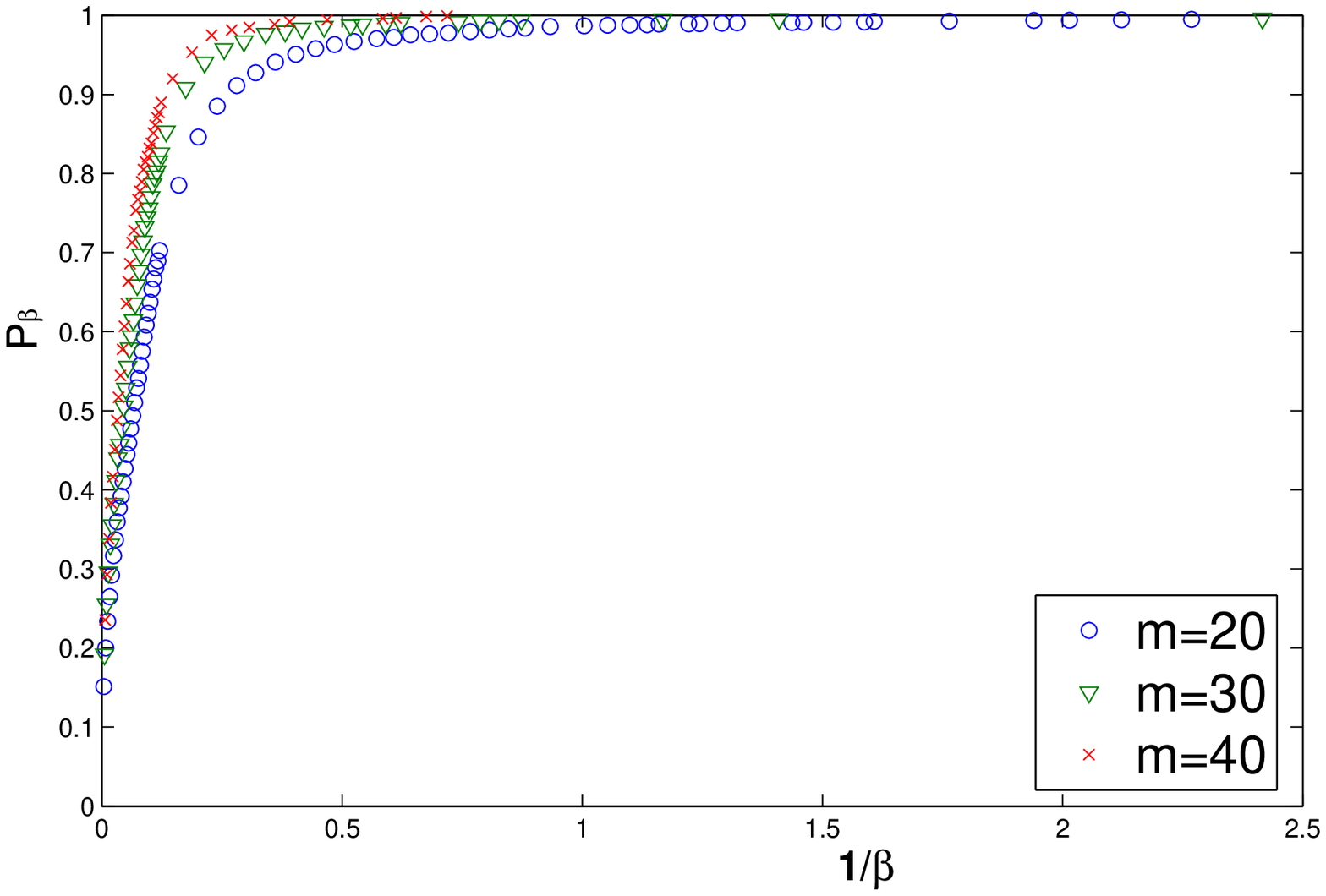}}
\caption{\label{2a}
\ensuremath{ {\cal P}_\beta }
is plotted as a function of \ensuremath{1/\beta} for the instances of 
Fig. \ref{1a}}
\end{figure}

\begin{figure}
\epsfxsize=10cm
\centerline{\epsffile[60 270 550 610]{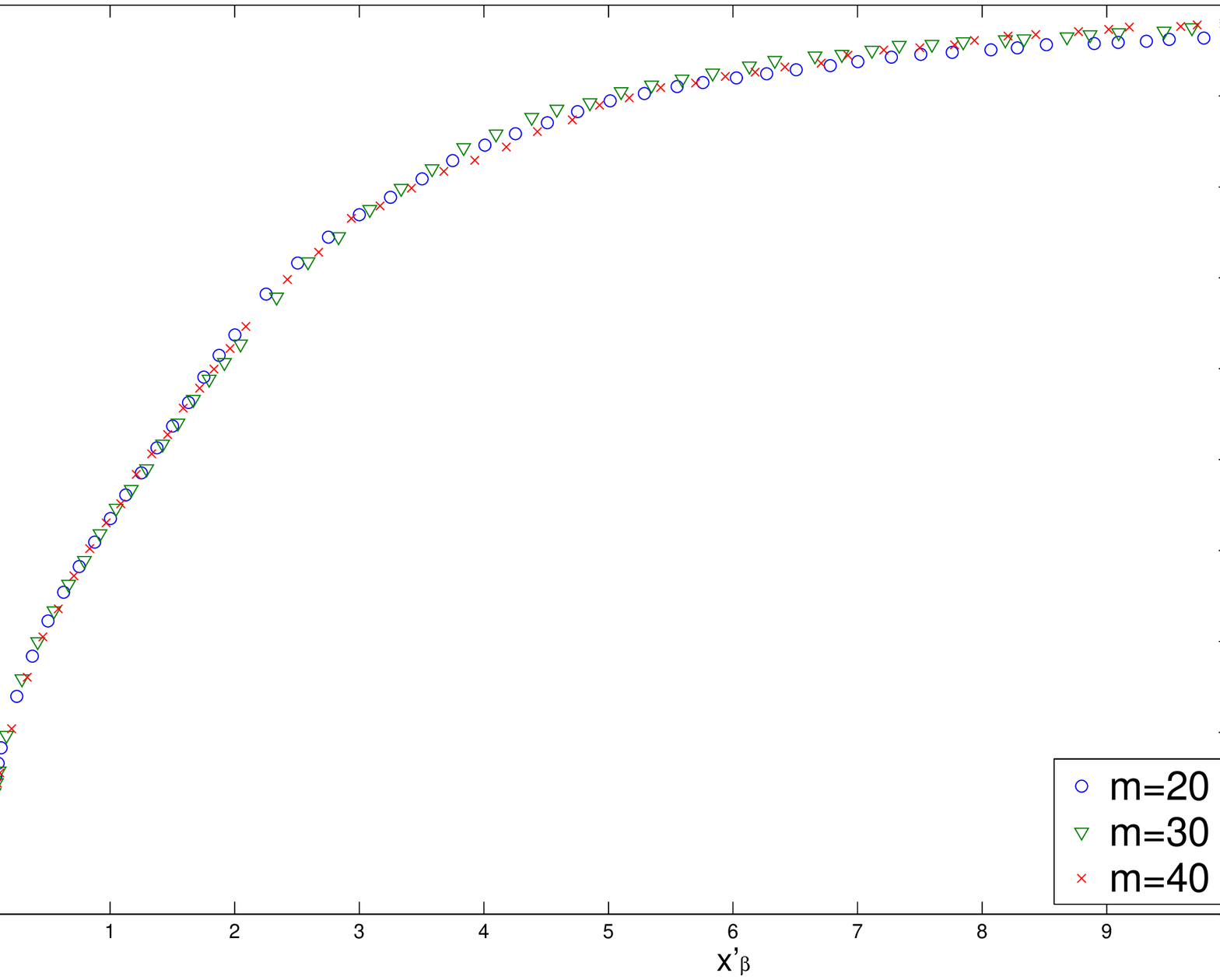}}
\caption{\label{2b}
\ensuremath{ {\cal P}_\beta }
is plotted as a function of \ensuremath{x'_\beta} for the instances of 
Fig. \ref{1a}.}
\end{figure}

\begin{figure}
\epsfxsize=10cm
\centerline{\epsffile[60 270 550 610]{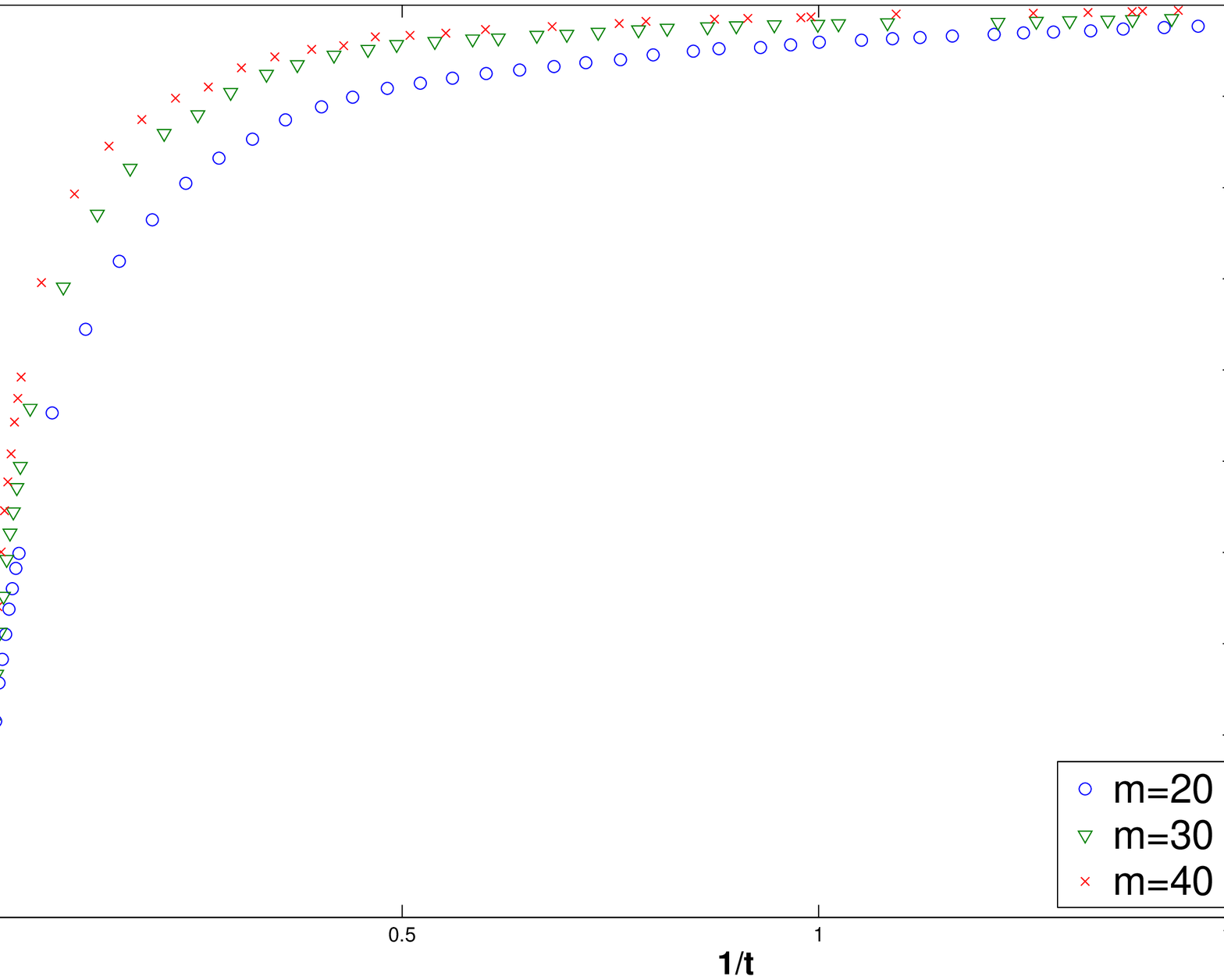}}
\caption{\label{3a}
\ensuremath{ {\cal P}_T}
is plotted as a function of \ensuremath{1/t} for the instances of 
Fig. \ref{1a}.}
\end{figure}

\begin{figure}
\epsfxsize=10cm
\centerline{\epsffile[60 270 550 610]{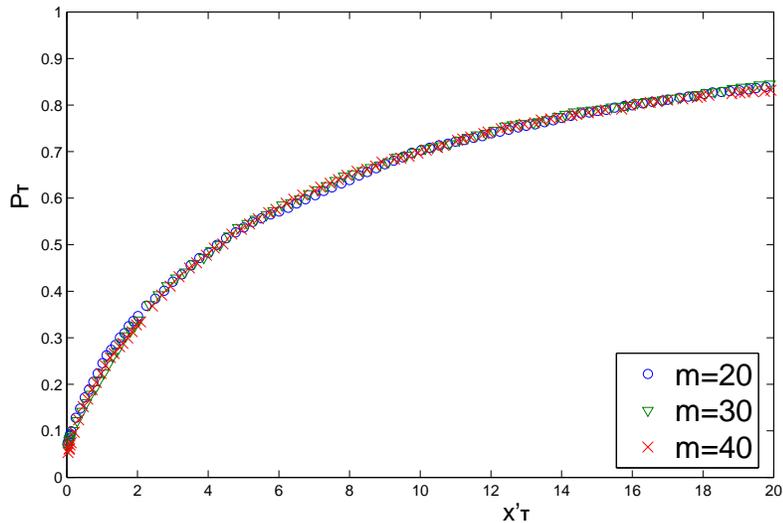}}
\caption{\label{3b}
\ensuremath{ {\cal P}_T}
is plotted as a function of \ensuremath{x'_T} for the instances of 
Fig. \ref{1a}.}
\end{figure}

In this case, the various elements of $A$, $b$ and $c$ take only the 
values $+1$ or $-1$ with probability $1/2$ each, namely,
\begin{equation}\label{bimodal}
P(y)=\frac{1}{2} [\delta(y-1)+\delta(y+1)]\,.                           
\end{equation}
The mean of this distribution is 0, and its variance is 1.
One problem associated with the discrete ensemble (\ref{bimodal}) is the 
finite probability to draw a degenerate LP problem. (Note that such 
degenerate, ill-defined  LP problems, comprize a set of zero measure in 
continuous ensembles such as the Gaussian ensemble.) In order to avoid these
degenerate solutions, we introduce a ``regularization" by which each matrix 
element $A_{ij}$, chosen at random from the ensemble (\ref{bimodal}), 
is multiplied by 
\begin{equation}\label{regularization}
f_{ij} = 1+[i+2(j-1)]\tilde\epsilon\,,                                 
\end{equation}
where $\tilde\epsilon $ is a small regularization parameter.  
(The other entries $b_i$ and $c_i$ take the values chosen at random with the 
probability (\ref{bimodal}) without any regularization.) In the numerical 
calculations we take $\tilde\epsilon = 10^{-5}$. (Note that the regularization
(\ref{regularization}) slightly splits the identical unregulated probability
distributions of the independent matrix elements $A_{ij}$. For small 
$\tilde\epsilon$ we expect this fact to have a negligible effect on the 
scaling behavior.)   
In Figs. \ref{1a} - \ref{3b} the distribution functions (\ref{pd}), 
(\ref{pb}), and (\ref{pt}) are plotted first as functions of the unscaled 
variables $\Delta$, $1/\beta$, and $1/t$, followed by the corresponding 
plots as functions of the scaled variables (\ref{xd'}), (\ref{xb'}), and 
(\ref{xt'}), respectively, for various values of $m$ while $n=2m$. Note that 
in terms of the scaled variables (\ref{xd'})-(\ref{xt'}), the data is
found to collapse to one distribution.

\pagebreak
\newpage

\subsection{The diluted bimodal distribution}

This is the distribution in which the random variable defined by 
(\ref{bimodal}) is replaced, for some values chosen at random, 
by 0. The resulting random variable is   
\begin{equation}
z = uy     
\label{z}   
\end{equation}
where $y$ is distributed according to (\ref{bimodal}) and $u$ is distributed 
according to
\begin{equation}
P(u) = p\delta(1-u)+(1-p)\delta(u)\,,
\label{pu1}
\end{equation}
where $0\leq p \leq 1$ is the dilution parameter. 
Thus, in our notations, $p=1$ corresponds to no dilution. 
The mean of the distribution of $z$ is $0$, and its variance is $p$.
We again apply the regularization process to the matrix $A$ by 
multiplying the matrix elements $A_{ij}$, chosen at random in this ensemble by 
(\ref{regularization}). 

\begin{figure}
\epsfxsize=8cm
\centerline{\epsffile[60 270 550 610]{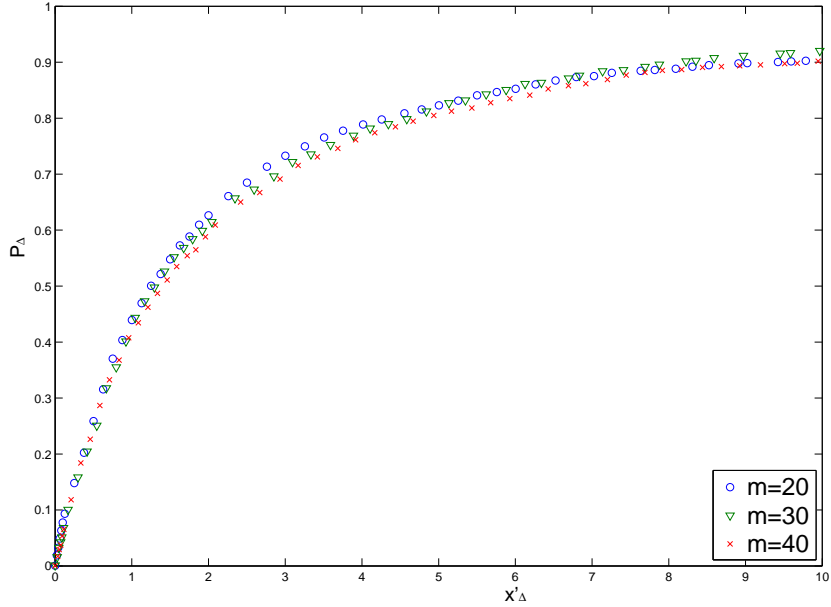}}
\caption{\label{4b}
\ensuremath{ {\cal P}_\Delta }
is plotted as a function of \ensuremath{x'_\Delta} for the diluted bimodal 
distribution, where $p=0.5$. As before, $n=2m$. 
The number of instances used in the simulation is 54951 for the $m=20$ case, 
41107 for the $m=30$ case and 50863 for the $m=40$ case. The number of 
converging instances for each case is 5000.}
\end{figure}

\begin{figure}
\epsfxsize=8cm
\centerline{\epsffile[60 270 550 610]{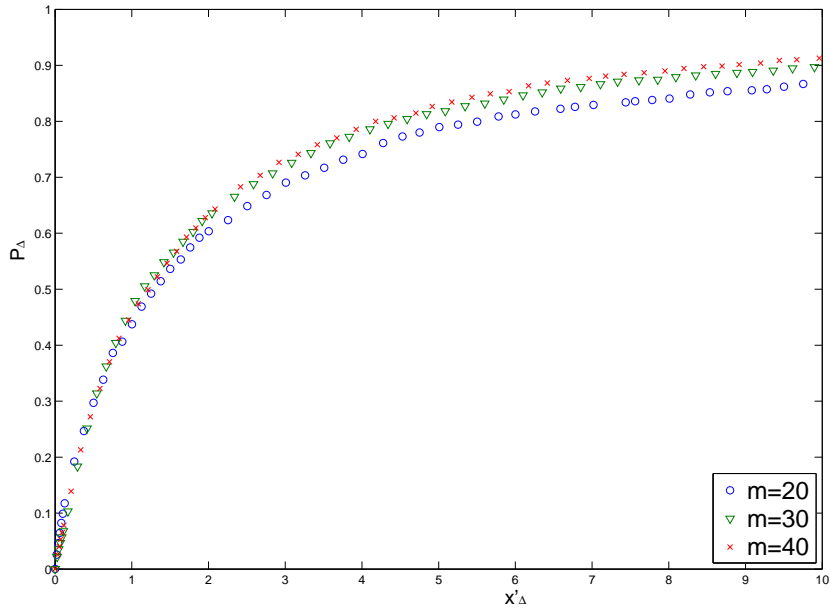}}
\caption{\label{7b}
\ensuremath{ {\cal P}_\Delta }
is plotted as a function of \ensuremath{x'_\Delta} for the diluted bimodal 
distribution, where $p=0.2$. As before, $n=2m$. 
The number of instances used in the simulation is 54620 for the $m=20$ case, 
37697 for the $m=30$ case, and 65367 for the $m=40$ case. The number of 
converging instances for these cases were, respectively, 4980, 3725 
and 5921.}
\end{figure}

The calculations were repeated for $p=0.5$  and $p=0.2$.  
In all cases a scaling function was found. 
The distributions $P_\Delta$ as a function of $x'_\Delta$ are 
presented in Fig. \ref{4b} for $p=0.5$, and in Fig. \ref{7b} for $p=0.2$. 
The analogous graphs of the distribution functions $P_\beta$ and $P_T$ for 
these diluted bimodal ensembles are presented in \cite{archive}.
Figs. \ref{4b} and \ref{7b}, as well as the corresponding figures for $P_\beta$
and $P_T$ indicate that the convergence in the limit $m\rightarrow\infty$ 
becomes slower as the dilution increases (i.e., as $p$ decreases).

\pagebreak

\newpage
\subsection{The uniform distribution}
\begin{figure}
\epsfxsize=8cm
\centerline{\epsffile[60 270 550 610]{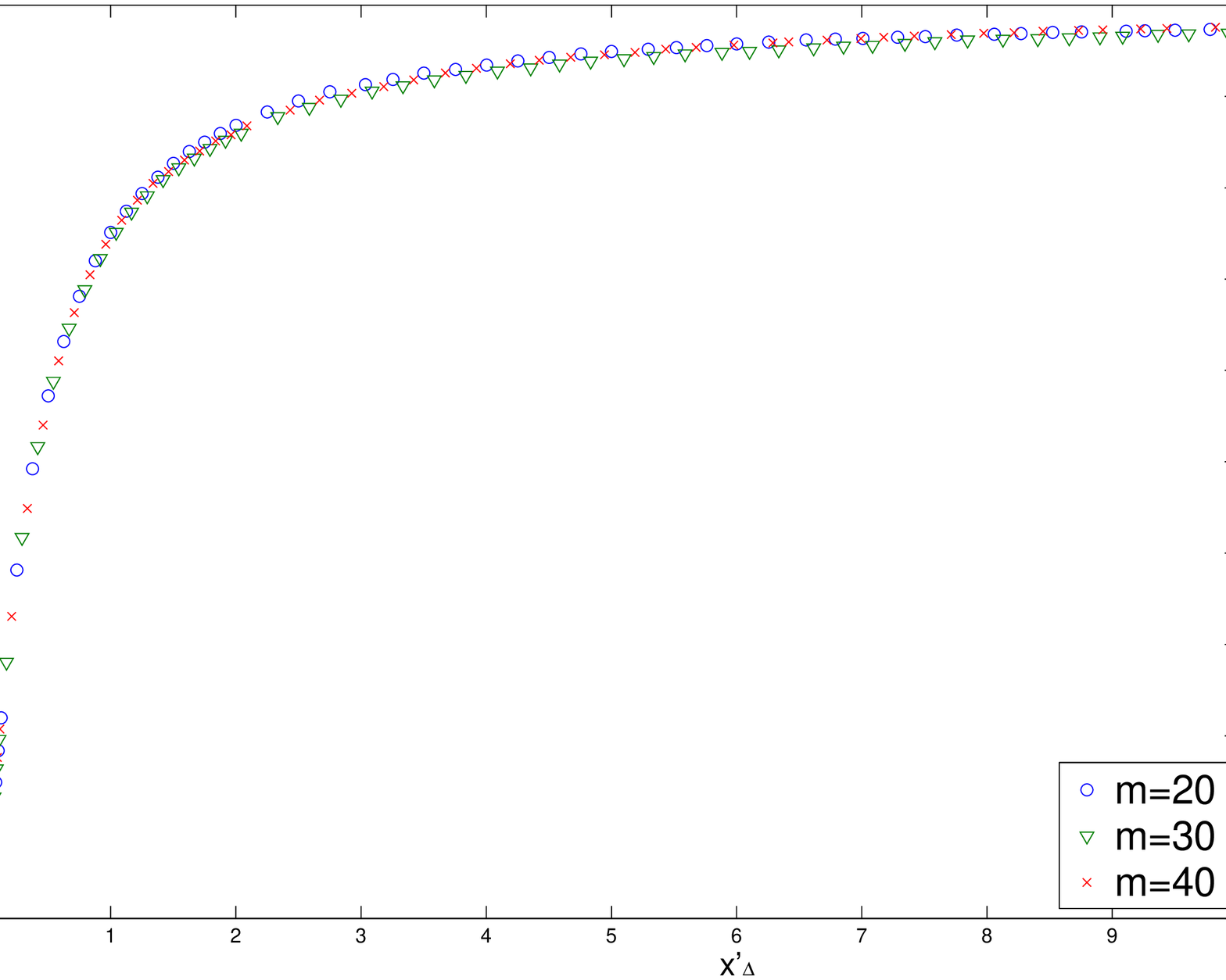}}
\caption{\label{11b}
\ensuremath{ {\cal P}_\Delta }
is plotted as a function of \ensuremath{x'_\Delta} for the uniform 
distribution. As before, \ensuremath{n=2m
}. The number of instances used in the simulation is 121939 for the 
\ensuremath{m=20} case, 91977 for the $m=30$ case and 112206 for the $m=40$ 
case. The number of converging instances for each case is 20000.}
\end{figure}

The distribution of the elements of $A$, $b$ and $c$ in this ensemble is
\begin{equation}\label{flat}
P(y) = \left\{\begin{array}{cc} 1\quad\quad & -1/2<y<1/2 \\{}&{}\\
0\quad\quad & {\rm otherwise}\end{array}\right.
\end{equation}
For this distribution the mean is 0 and the variance is 
$ {1\over 12}$. 
The distribution function (\ref{pd}) is presented in terms of the scaling 
variable (\ref{xd'}) in Fig. \ref{11b}. All the other distributions
related to this ensemble are presented elsewhere \cite{archive}.

\pagebreak

\newpage
\section{Universality}
In the previous section we demonstrated that for large $m$ and $n$, the 
distribution functions (\ref{pd}) - (\ref{pt}) depend on $\Delta,\beta, T$ 
and on $m$ and $n$ only via the scaling variables (\ref{xd'}) - (\ref{xt'}).
A natural question which arises is whether the distribution functions 
$P_\Delta$, $P_\beta$ and $P_T$ are universal \cite{bffs, pla}. 
In other words, we ask whether all probability ensembles of LP problems, or 
at least a large family thereof, yield the same functions $P_\Delta$, 
$P_\beta$ and $P_T$ of the scaling variables
\begin{equation}
x_\Delta = a^{(\mu)}_\Delta (n/m)x'_\Delta,
\label{xdd}
\end{equation}
\begin{equation}
x_\beta = a^{(\mu)}_\beta (n/m)x'_\beta
\label{xb} 
\end{equation}
and
\begin{equation}
x_T = a^{(\mu)}_T (n/m)x'_T
\label{xt}
\end{equation}
where $x'_\Delta$, $x'_\beta$ and $x'_T$ are defined by 
(\ref{xd'}),(\ref{xb'}) and (\ref{xt'}), and the scale factors are the
generalizations of $a^{(g)}_\Delta$, $a^{(g)}_\beta$ 
and $a^{(g)}_T$ of (\ref{ad}), (\ref{xbgauss}) and (\ref{xTgauss})? 

\begin{figure}
\epsfxsize=12cm
\centerline{\epsffile[60 270 550 610]{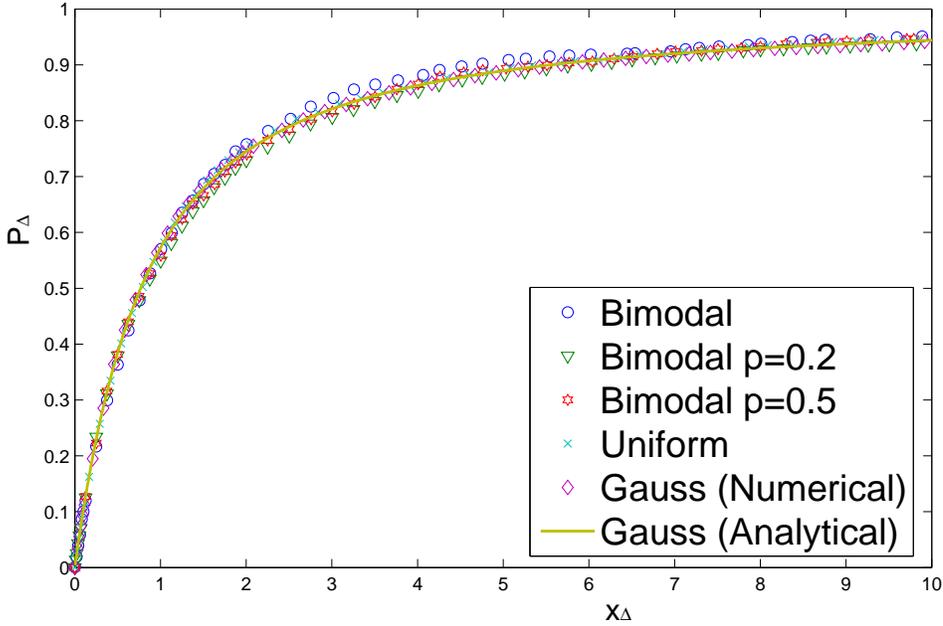}}
\caption{\label{14}
$P_\Delta$ as a function of $x_\Delta$ (for $n=2m$ and $m=40$). The graphs 
are scaled to fit the theoretical Gaussian result by appropriate choice of
the factors $a^{(\mu)}_\Delta$.}
\end{figure}

For the Gaussian distribution we denote $\mu=g$, for 
the undiluted bimodal distribution $\mu=p=1$, for 
the various diluted bimodal distributions $\mu=p:=0.5, 0.2$, and for the 
uniform distribution $\mu=u$. Throughout this paper we take $n=2m$.
The distributions of the $\Delta_{min}$ for $m=40$ for the various ensembles 
are presented in Fig. \ref{14}.

The scale factors $ a^{(\mu)}_\Delta$ are 
chosen so as to minimize the deviation of the specific distribution 
$P_\Delta$ from the Gaussian distribution ${\cal F}(x_\Delta)$ of 
(\ref{scaling.delta}), that was calculated analytically in \cite{bffs, pla}. 
This is done, as usual, by least squares fit.
The Gaussian distribution was taken with variance $\sigma_{(g)}^2=1$ 
in our calculation. For the Gaussian distribution the numerical results were 
found to fit the analytical result (\ref{scaling.delta}) with 
$ a^{(g)}_\Delta \simeq 0.564\simeq  {1\over \sqrt{\pi}},$ as expected from 
(\ref{ad}) for $n=2m$. For the bimodal distribution 
it was found that $ a^{(1)}_\Delta \simeq 0.558 \simeq {0.989\over 
\sqrt{\pi}}$.  
For the diluted bimodal distributions at $p=0.5$ we found $ a^{(0.5)}_\Delta 
\simeq  0.581 \simeq {1.030\over \sqrt{\pi}}$, and for 
the more diluted ensemble at $p=0.2$ we found 
$ a^{(0.2)}_\Delta \simeq  0.687 \simeq {1.218\over \sqrt{\pi}}$. Finally, 
for the uniform distribution we obtained $ a^{(u)}_\Delta \simeq
1.957  \simeq {3.469 \over{\sqrt{\pi}}} \simeq \sqrt{12.030\over \pi}\,.$ 
Recall that the variances of these distributions are $\sigma_{(1)}^2=1$, 
and $\sigma_{(u)}^2={1\over 12}$, respectively.
Therefore, on the basis of these numerical results we conjecture that 
\beq\label{conjectureDelta}
\sigma_{(g)}a^{(g)}_\Delta =  \sigma_{(1)}a^{(1)}_\Delta = 
\sigma_{(u)}a^{(u)}_\Delta\,,
\eeq
and that their common value is given by (\ref{ad}), that is equal to 
$1/\sqrt{\pi}$ for $n=2m$.

Our numerical results indicate that the scale factors  
$ a^{(p)}_\Delta$ of the diluted distributions deviate from this 
simple law, and this deviation seems to be more pronounced for higher 
dilution (smaller $p$). For these distributions $\sigma_{(p)}^2 = p$, leading 
to $ a^{(0.5)}_\Delta \sigma_{(0.5)} \simeq 0.411$ and 
$ a^{(0.2)}_\Delta \sigma_{(0.2)} \simeq 0.307$, which differ significantly
from the more or less common value of this product for the undiluted 
distributions, namely $1/\sqrt{\pi}\simeq 0.564$. 

From Fig. \ref{14} we see that the distribution functions of the scaling
variables $x_\Delta$ corresponding to the convergence rates $\Delta$ approach 
a universal function, that is identical to the one that is found analytically
for the Gaussian distribution, and is given by (\ref{scaling.delta}). For the
undiluted distributions also the scale factors were found to agree with  
(\ref{ad}).

The proportionality of $a^{(\mu)}_\Delta $ to $1/\sigma_{(\mu)}$ probably
results from the fact that if all parameters $c_i$ and $A_{ij}$ in 
(\ref{standard}) are rescaled by some common factor, also the $\Delta_i$,
defined by (\ref{deltas}), are rescaled by the same factor.
For the diluted distributions, the behavior of the scaling factors 
is different. Behavior of similar nature is found also for the distribution
functions $P_\beta$ and $P_T$. 

\begin{figure}
\epsfxsize=9cm
\centerline{\epsffile[60 270 550 610]{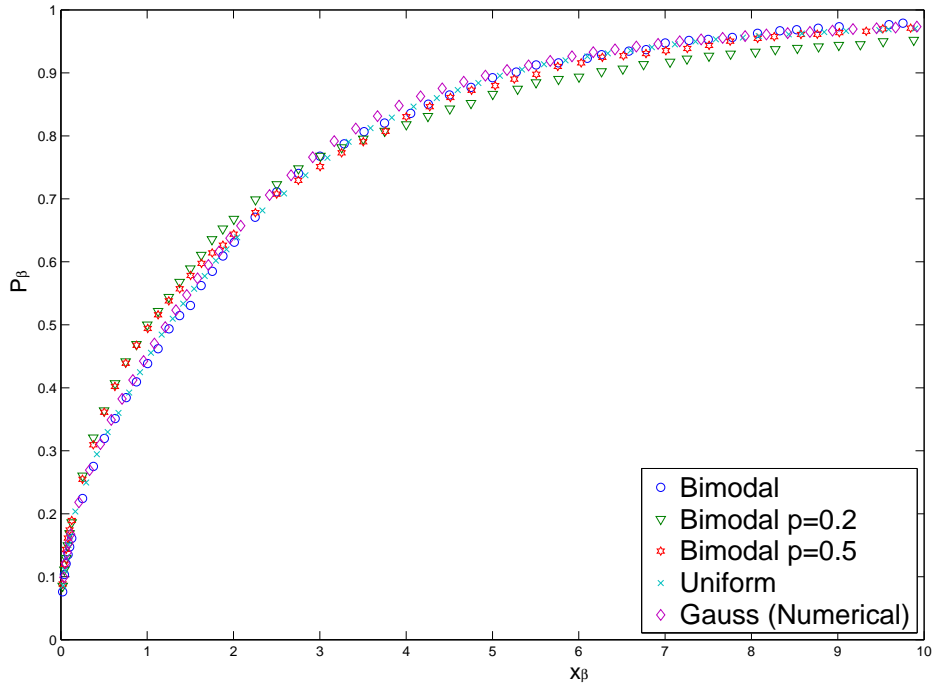}}
\caption{\label{15}
$P_\beta$ as a function of $x_\beta$ for all the distributions checked 
(for $n=2m$ and $m=40$), where the scale factors $a^{(\mu)}_\beta$ were 
found by least squares fit to the distribution for the Gaussian ensemble, 
which was found numerically as well.}
\end{figure}

The distributions $P_\beta$ in terms of $x_\beta$ are presented in 
Fig. \ref{15} for $m=40$. Using the scale factor $a^{(g)}_\beta = 1$, we 
find numerically, by least square fit to the case of the Gaussian distribution
(that was also found numerically), the other scale factors to be 
$a^{(1)}_\beta \simeq 0.952$, $a^{(0.5)}_\beta \simeq 1.189$, 
$a^{(0.2)}_\beta \simeq 1.943$ and $a^{(u)}_\beta \simeq 0.960. $ 

The scale factors $a^{(g)}_\beta, a^{(1)}_\beta$ and $a^{(u)}_\beta$ are 
pretty close to each other (and to unity), while $a^{(0.5)}_\beta$ and 
$a^{(0.2)}_\beta$ deviate from them significantly. 

We note from (\ref{barrier}) that the barriers $\beta_i$ are 
invariant under global rescaling of all the matrix elements $A_{ij}$ by the
same factor. Thus, as long as the optimal vertex $x_B^*$ does not have too 
many anomalously small or large components, $P_\beta$ is expected to be 
independent of the variance, and the numerical values we obtained for 
$a^{(1)}_\beta \simeq a^{(u)}_\beta \simeq a^{(g)}_\beta$ seem to support 
this expectation for the undiluted ensembles. The scale factors
$a^{(0.5)}_\beta, a^{(0.2)}_\beta$ deviate from that common value, with 
$a^{(0.2)}_\beta$ deviating very significantly. 

Our numerical results, displayed in Fig. \ref{15}, show that the 
distributions $P_\beta (x_\beta)$ found for the diluted ensembles deviate 
from the ones found for the undiluted ensembles. 
Therefore, for the diluted ensembles, although we have good indication for 
universality, this point clearly requiers further research.

\begin{figure}
\epsfxsize=9cm
\centerline{\epsffile[60 270 550 610]{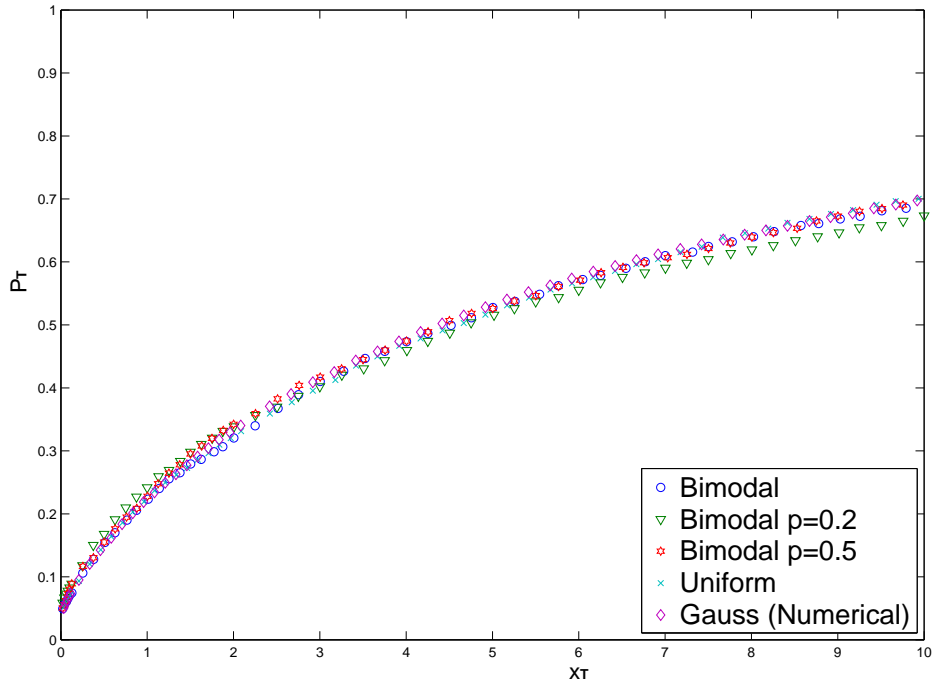}}
\caption{\label{16}
$P_T$ as a function of $x_T$ for all the distributions checked (for $n=2m$ 
and $m=40$). The scale factors $a^{(\mu)}_T$ were 
found by least squares fit to the distribution for the Gaussian ensemble, 
which was found numerically as well.}
\end{figure}

The distributions $P_T$ in terms of $x_T$ are presented in Fig. \ref{16}, 
for $m=40$. 

Using the scale factor $a^{(g)}_T = 1$ we find
numerically $a^{(1)}_T \simeq 1.008$,  
$a^{(0.5)}_T \simeq 1.240$, $a^{(0.2)}_T \simeq  2.088$, 
and $a^{(u)}_T \simeq 3.399 \simeq \sqrt{11.553}\,.$
From (\ref{T}) we see that under global rescaling of the parameters 
$A_{ij}, b_i, c_i$ in (\ref{standard}), the scaled variables $x'_\Delta$ and 
$x'_T$ are rescaled in the same way. Thus, since $a^{(g)}_\Delta $ is 
proportional to $ 1/\sigma $, also $a^{(g)}_T$ should satisfy a similar
proportionality. Since for the Gaussian distribution we have taken 
$\sigma_{(g)}^2 = 1$ and $a^{(g)}_T = 1\,,$ our numerical results, where we 
find $a^{(u)}_T \simeq \sqrt{12}\,,$ suggest that for the undiluted ensembles
\beq\label{Tconjecture}
\sigma_{(g)}a^{(g)}_T =  \sigma_{(1)}a^{(1)}_T = \sigma_{(u)}a^{(u)}_T\,,
\eeq
taking the value unity in our case. For the diluted ensembles 
we find $ \sigma_{(0.5)} a^{(0.5)}_T \simeq 0.877$ and 
$\sigma_{(0.2)} a^{(0.2)}_T \simeq 0.933$, deviating significantly from unity. 
For $p=0.2$ a significant deviation of $P_T$ from the other distributions is 
found. 

\section{Summary and Discussion}
In this paper we have presented ample numerical evidence for the fact that 
the asymptotic distribution functions $P_\Delta, P_\beta$ and $P_T$
are scaling functions. 

In particular, all the {\em undiluted} ensembles of LP 
problems which we studied, seem to have the {\em same} set of asymptotic 
distribution functions, when the latter are expressed in terms of the 
scaling variables $x_\Delta, x_\beta$ and $x_T$. 
Furthermore, we have found that then the following combinations of scaling 
factors, $\sigma_{(\mu)}a^{(\mu)}_\Delta, 
a^{(\mu)}_\beta$ and $\sigma_{(\mu)}a^{(\mu)}_T$ are independent of the 
probability distribution. Therefore, in particular, $a^{(\mu)}_\Delta$ should 
satisfy (\ref{ad}) that was found for the Gaussian distribution, 
while $ a^{(\mu)}_\beta$ and $a^{(\mu)}_T$, as well as corresponding 
distributions, are yet to be found analytically.

Based the results presented in this paper, as well as the results of 
\cite{bffs, pla}, we conjecture the scaling behavior of the various 
undiluted distribution functions (and the corresponding scaling factors) is 
{\em universal}, i.e., that it is robust and should be valid in a large 
class of ensembles of LP problems, in which the $(A,b,c)$ data are taken from 
a distribution with zero mean and finite variance.

We have also studied the effect of dilution, namely, imposing that the 
parameters $A_{ij}, c_i$ and $b_i$ in (\ref{standard}) could vanish with 
finite probability $1-p$. Specifically, we have studied
the effects of dilution only on the bimodal distribution.

Our findings, depicted in Figs. \ref{4b} and \ref{7b}, as well as the 
corresponding figures for $P_\beta$ and $P_T$, indicate that the 
convergence in the limit $m\rightarrow\infty$ becomes slower as the dilution 
increases (i.e., as $p$ decreases). More importantly, our numerical 
results {\em indicate} that the diluted distributions may exhibit scaling 
behavior as well, but with scaling factors which are different from those 
of the undiluted ensembles which belong in the universality class of the 
undiluted Gaussian ensemble. 
Moreover, the corresponding asymptotic scaling distribution functions 
$P_\Delta, P_\beta$ and $P_T$ of these diluted ensembles deviate 
sometimes from those of the corresponding ones of the undiluted Gaussian 
universality class. This deviation appears to become more pronounced as 
dilution increases, and it may be related to the fact that a finite fraction 
of the admissible LP instances in diluted ensemble may have an optimal vertex 
$x_B^*$ which has too many anomalously small or large components. 

Generally speaking, dilution seems to have interesting effects, which are not 
completely understood, and call for further investigation. Specific
questions are motivated by the present work. In particular, are
the asymptotic scaling distribution functions 
$P_\Delta, P_\beta$ and $P_T$, which we computed numerically, really different 
from the ones found for the undiluted Gaussian ensemble (with possibly scale 
factors which differ from the Gaussian ones), or the deviations of 
these functions, indicated by our numerical results, are merely effects of 
the slower convergence towards asymptotics? If they are different - do they 
form another universality class?

Finally, we would like to raise a question which may be of practical 
importance. Thus, imagine that all LP problems used in practice 
(or at least, a large fraction thereof) are collected into 
an unbiased probability ensemble.  
How is this distribution of {\em realistic} LP problems related to the 
ensembles studied in this paper? Does it really relate to a universality 
class (or classes) of ensembles of LP problems studied here? Does
it agree more with the diluted or the undiluted ensembles? These questions 
clearly pose important conceptual challenges for further investigation, and 
also have practical implications.

\newpage

Acknowledgements:
It is a great pleasure to thank our colleagues Asa Ben-Hur and Hava Siegelmann
for very useful advice and discussions. This research was supported in part 
by the Shlomo Kaplansky Academic Chair, by the Technion-Haifa University 
Collaboration Fund, by the US-Israel Binational Science Foundation (BSF), 
by the Israeli Science Foundation (ISF), and by the Minerva Center of 
Nonlinear Physics of Complex Systems.


\begin{thebibliography}{10}

\bibitem{Hertznn-optimwang}
J.~Hertz, A.~Krogh, and R.~Palmer.
\newblock {\em Introduction to the Theory of Neural Computation}.
\newblock Addison-Wesley, Redwood City, 1991.


\bibitem{mead}
C.~Mead.
\newblock {\em Analog VLSI and Neural Systems}.
\newblock Addison-Wesley, 1989.


\bibitem{bio}
D.~Bray. {\em Nature} {\bf 376}, 307 (1995);~~ A. ~Ben-Hur and H.T.~ Siegelmann. {\em Proceedings of MCU 2001},
{\em Lecture Notes in Computer Science} {\bf 2055}, pages 11-24, M. ~Margenstern and Y. ~Rogozhin (Editors),
Springer Verlag, Berlin 2001 (and references therein).


\bibitem{ott} E. Ott, {\em Chaos in Dynamical Systems}. Cambridge University
Press, Cambridge, England, 1993.


\bibitem{brockett}
R.~W. Brockett.
\newblock {\em Linear Algebra and Its Applications} {\bf 146}, 79 (1991);~~
M.S. Branicky. \newblock Analog computation with continuous {ODE}s.
\newblock In {\em Proceedings of the IEEE Workshop on Physics and Computation
}, pages 265--274, Dallas, TX, 1994.


\bibitem{faybusovich}
L.~Faybusovich.
\newblock {\em IMA Journal of Mathematical Control and Information}
{\bf 8}, 135 (1991).

\bibitem{helmke-moore}
U.~Helmke and J.B. Moore.
\newblock {\em Optimization and Dynamical Systems}.
\newblock Springer Verlag, London, 1994.

\bibitem{Papadimitriou}
C.~Papadimitriou.
\newblock {\em Computational Complexity}.
\newblock Addison-Wesley, Reading, Mass., 1995.

\bibitem{rmt}
M.L.~Mehta.
\newblock {\em Random Matrices} (2nd ed.).
\newblock Academic Press, San-Diego, CA, 1991.


\bibitem{nucl}
T.A.~Brody, J.~Flores, J.B.~French, P.A.~Mello, A.~Pandey and S.S.M. Wong.
\newblock {\em Rev. Mod. Phys.} {\bf 53}, 385 (1981).


\bibitem{chaos}
O.~Bohigas, M.-J.~Giannoni and C. Schmit.
\newblock {\em Phys. Rev. Lett.} {\bf 52}, 1 (1984);~~O.~Bohigas.
\newblock Random Matrix Theories and Chaotic Dynamics.
\newblock In {\em Chaos and Quantum Physics}, Proceedings of the Les-Houches
Summer School, Session LII, 1989, M.-J.~Giannoni, A.~Voros and
J.~Zinn-Justin, (eds.), North-Holland, Amsterdam, The Netherlands, 1991.

\bibitem{bffs}
A.~Ben-Hur, J.~Feinberg S.~Fishman, and H.T.~Siegelmann, \newblock {\em J.~ 
Complexity} {\bf 19}, 474 (2003) \newblock (preprint cs.CC/0110056). 

\bibitem{pla}
A.~Ben-Hur, J.~Feinberg S.~Fishman, and H.T.~Siegelmann, \newblock {\em Phys.~ 
Lett.~A} {\bf 323}, 204 (2004) \newblock (preprint cond-mat/0110655). 

\bibitem{dds}
H.T. Siegelmann, A. Ben-Hur and S. Fishman,
\newblock {\em Phys. Rev. Lett.} {\bf 83}, 1463 (1999);
A.~Ben-Hur, H.T. Siegelmann, and S.~Fishman. \newblock {\em J.~ 
Complexity} {\bf 18}, 51 (2002). 


\bibitem{SF}
H.T. Siegelmann and S.~Fishman.
\newblock {\em Physica} {\bf 120}, 214 (1998).


\bibitem{toda}
L.~Faybusovich.
\newblock {\em Physica} {\bf D53},  217 (1991).


\bibitem{BCSS}
L.~Blum, F.~Cucker, M.~Shub, and S.~Smale.
\newblock {\em Complexity and real Computation}.
\newblock Springer-Verlag, 1999.


\bibitem{traub}
S.~Smale. \newblock {\em Math. Programming} {\bf 27}, 241 (1983);
M.J. Todd.
\newblock {\em Mathematics of Operations Research} {\bf 16}, 671 (1991).


\bibitem{lpsmaleTodd-models}
S.~Smale. \newblock {\em Math. Programming} {\bf 27}, 241 (1983).\\
M.J. Todd.
\newblock {\em Mathematics of Operations Research} {\bf 16}, 671 (1991).

\bibitem{shamir}
R.~Shamir.
\newblock {\em Management Science} {\bf 33}(3), 301 (1987).



\bibitem{Ye-book}
Y.~Ye.
\newblock {\em Interior Point Algorithms: Theory and Analysis}.
\newblock John Wiley and Sons Inc., 1997.


\bibitem{reif} V.~Y.~Pan and J.~Reif,
\newblock {\em Computers and Mathematics (with Applications)} {\bf 17}, 1481 
(1989). 


\bibitem{wilson}
K.G. Wilson and J. Kogut, {\it Phys.~Rep.} {\bf 12}, 75 (1974).

\bibitem{anderson}
E. Abrahams, P. W. Anderson, D. C. Licciardelo and T. V. Ramakrishnan,
{\it Phys. ~Rev. ~Lett.} {\bf 42} 673 (1979).


\bibitem{universality}
Some papers that treat random {\em real} rectangular matrices, such as the
matrices relevant for this work (which are not necessarily Gaussian ), are:
A.~ Anderson, R.~C. Myers and V.~ Periwal, {\it Phys. Lett.}{\bf B 254}, 89
(1991); ~~ {\it Nucl. ~Phys.} {\bf B 360}, (1991) 463 (Section 3);~~
J.~Feinberg and A.~Zee, {\it J. Stat. Mech.} {\bf 87}, 473
(1997);~~For earlier work see: G.M. Cicuta, L. Molinari, E. Montaldi
and F. Riva, {\it J. Math.Phys.} {\bf 28}, 1716 (1987).


\bibitem{archive}
Y. S. Avizrats, J. Feinberg and S. Fishman, {\sl Scaling and Universality of 
the Complexity of Analog Computation}, to appear.



\end{thebibliography}
\end{document}